\documentclass[aps,showpacs,floatfix,twocolumn]{revtex4}
\usepackage{amsmath}
\usepackage{graphicx}
\begin{document}

\title{Field-induced layer thinning transition on free-standing smectic films}

\author{Maria S. S. Pereira, Marcelo L. Lyra, and  Italo N. de Oliveira \email{italo@if.ufal.br}}

\affiliation{
Instituto de F\'{\i}sica, Universidade Federal de Alagoas 
57072-970 Macei\'o-AL, Brazil}
\date{\today}

\begin{abstract}
Strongly anchored free-standing smectic films usually present a stepwise reduction of the number of layers when the temperature is raised above the smectic-isotropic bulk transition temperature. Here, we demonstrate that a field-induced layer thinning transition can take place in smectic films with a negative dielectric anisotropy even below the bulk transition temperature. Using an extended McMillan's model, we provide the phase diagram of this layering transition and show that, when the field is raised above the bulk transition field, the film thickness reduction is well described by a power-law with an exponent that depends on the temperature and the aspect ratio of the liquid crystal molecule.
\end{abstract}

\pacs{61.30.Hn, 64.70.mf, 61.30.Gd}
\maketitle

Surface and external field effects on the liquid crystal properties
have attracted an appreciable interest over the past decade, being
the subject of several theoretical and experimental works \cite{bahr,chao,barbero,sprunt,lelidis,rosenblatt}.
In particular, these systems exhibit a rich phenomenology associated with 
the surface anchoring, as well as with the coupling between orientational order and 
electric or magnetic fields \cite{degennes}. In fact, surface ordering and 
field-induced reorientation are essential mechanisms to understand a 
great variety of problems involving liquid crystals in different areas, 
such as phase transitions, nonlinear optics, and colloidal dispersions.

It is well known that surface effects may stabilize the smectic 
order in free-standing films well above the bulk transition 
temperature \cite{bahr1}. Depending on the strength of the surface
anchoring, several unusual phenomena may be observed
in such systems. A prominent example of surface-induced phenomenon 
is the layer thinning transition in free-standing smectic films
which consists in a stepwise reduction of the film thickness as the 
temperature is raised above of the bulk transition temperature 
\cite{stoebe}. By using the optical reflectivity technique, experimental
works have observed the layer thinning transition in a large variety of 
compounds \cite{stoebe,mol,demikhov,jin,johnson,pankratz,picano}. In all cases,
the film thickness reduction was suitably described by a simple power-law 
expression $N(t) \propto t^{-\nu}$, where $N$ is the number of layers and 
$t$ is the reduced temperature. However, it was noticed that the exponent $\nu$ 
assumes values in the range  $0.52 \leq \nu \leq 0.82$ for different compounds.
Recent theoretical investigations have demonstrated that the main
experimental characteristics of the thinning transitions can be 
understood in the light of mean-field models \cite{mirantsev,sullivan,canabarro}. 
In particular, the experimental range of the thinning exponent
was reproduced by varying some typical parameters of the models, as the
surface anchoring \cite{sullivan} and the interaction strength for the
smectic-A phase \cite{canabarro}.

Recently, the effects of an external field on the liquid crystal 
phase transitions have been systematically investigated. In particular, 
experimental studies have reported that an external field may promote 
the emergence of ordered phases well above the isotropic-nematic and 
isotropic-smectic bulk transition temperatures \cite{sprunt,lelidis1}. Close to 
the nematic - smectic A transition, birefringence measurements revealed that a 
strong electric field suppresses the nematic fluctuations in systems with a 
positive dielectric anisotropy, resulting in a field-driven crossover from first 
to second order phase transition \cite{lelidis}. An inverse crossover
has also been  identified in the Fr\'{e}edericksz transition in thin
homeotropic cells of a liquid crystal with a negative dielectric 
anisotropy \cite{rosenblatt}. In free-standing smectic films, an 
optical field was observed to induce a layer thickening in 
photosensitive samples \cite{dolganov2}. Further, theoretical
investigations have predicted that a magnetic field may affect
the layering 
transition, enhancing the transition temperature of films with a positive diamagnetic susceptibility anisotropy \cite{mirantsev2}. However, the possibility
of a controlled reduction of the film thickness by an external
field has not been explored so far. Such phenomenon would establish 
a new theoretical and experimental ground to study the joint surface and field effects
on the liquid crystal order and its dimensional reduction.

In the present letter, we demonstrate that a field-induced layer thinning transition
can indeed take place in free-standing smectic films even below to smectic-isotropic bulk transition temperature. 
We will use an extended McMillan model to explicitly take into account the discrete layered structure 
and the surface anchoring energy of thin smectic films. For strongly anchored films with a negative dielectric anisotropy, 
we will show that the external electric field induces a power-law stepwise reduction of the film 
thickness similar to the standard temperature-induced thinning transition. 
This phenomenon contrasts with the field-enhanced order predicted to occur in films with positive anisotropy\cite{mirantsev2}.
Further, we will characterize the dependence of the effective power-law exponent with the temperature and the aspect ratio of the liquid crystal molecule.  

A free-standing smectic film is described as a stack of smectic layers confined by a 
surrounding gas \cite{degennes}. Due to strong surface interactions, the
molecular alignment tends to be normal to the layer's plane and the film 
can be considered as a smectic monodomain. Nevertheless, an electric
field perpendicular to the layer's plane may induce a molecular reorientation 
in systems with a negative dielectric anisotropy. In a mean-field approach for a film with $N$ discrete 
layers, the effective potential felt by a molecule located at the $i$-th smectic layer can be written as \cite{mirantsev}:
\begin{eqnarray}
V_1(z_1,\theta_1)= &-&\frac{V_0}{3}\left[s_1+s_2+ 3W_0/V_0 + \varepsilon^*_aE^2/ V_0\right. \nonumber \\ 
&+& \left. \alpha \cos(2\pi z_1/d)\left(\sigma_1+\sigma_2\right)\right]P_2\left(\cos \theta_1\right)~~~
\end{eqnarray}
\begin{eqnarray}
V_i(z_i,\theta_i)&=& -\frac{V_0}{3}\left[\sum_{j=i-1}^{i+1}s_j + \varepsilon^*_aE^2/V_0\right. \nonumber \\
&+&\left.\alpha \cos(2\pi z_i/d)\left(\sum_{j=i-1}^{i+1}\sigma_j\right)\right]P_2\left(\cos \theta_i\right) 
\end{eqnarray}
\begin{eqnarray}
V_N(z_N,\theta_N)&=& -\frac{V_0}{3}\left[s_N+s_{N-1}+ 3W_0/V_0 + \varepsilon^*_aE^2/V_0\right. \nonumber \\
&+& \left.\alpha \cos(2\pi z_N/d)\left(\sigma_N+\sigma_{N-1}\right)\right]P_2\left(\cos
\theta_N\right).
\end{eqnarray}
Here, $P_2(cos \theta_i)$ is the second-order Legendre polynomial
with $\theta_i$ being the angle between the 
long axis of a molecule at the $i$-th layer and the $z$ direction. $s_i$ and $\sigma_i$
are the orientational and translational order parameters in the $i$-th layer,
respectively. $V_0$ is a parameter of the microscopic model that 
determines the scale of the nematic-isotropic transition temperature \cite{mcmillan}. 
The parameter $\alpha$  is related to the length of alkyl chains of calamitic 
molecules through the expression $\alpha =2 exp[-(\pi r_0 /d)^2]$, 
where $r_0$ is a characteristic length associated to the length of the molecular 
rigid section, and $d$ is the smectic layer spacing. The parameter $W_0$ 
corresponds to the strength of the homeotropic surface anchoring which is assumed 
to be short ranged. $E$ represents an external electric field which is applied
perpendicularly to the layer plane. $\varepsilon^*_a$ is defined as 
$\varepsilon^*_a = \varepsilon_a/4\pi n_0$, where $\varepsilon_a$ is
the dielectric anisotropy of the system and $n_0$ is the density of 
particles. In liquid crystal samples with positive dielectric anisotropy, an external electric field tends to reinforce the nematic and smectic order parameters.
However, in compounds with negative anisotropy $\varepsilon_a$, a perpendicular electric field  reduces 
the smectic order through the reorientation of the molecular alignment. It is this scenario that we will
explore in what follows, concerning the possibility of field-controlled reduction of the film thickness.

The local order parameters $s_i$ and $\sigma_i$ satisfy the 
self-consistent equations:
\begin{equation}
s_i = \langle P_2 (cos \theta_i) \rangle_i
\end{equation}
\noindent and
\begin{equation}
\sigma_i = \langle P_2 (cos \theta_i)cos(2\pi z_i/d) \rangle_i,
\end{equation}

\noindent with the thermodynamical averages being computed from the one particle
distribution function in the $i$-th smectic layer, given by
\begin{equation}
 f_i(z_i,\theta_i) \propto exp \left[-V_i/k_BT\right]~~~,
\end{equation}
where $k_B$ is the Boltzmann constant and $T$ is the temperature. 
The solutions of Equations (4) and (5) minimize the total Helmholtz free-energy \cite{canabarro}.
 In the absence of an external
electric field, this model predicts a similar McMillan's phase diagram
for the internal layers when the film thickness $l$
is much larger than the surface penetration length $\delta$ \cite{selinger}. 
In particular, we will restrict the present work to this limit ($l >> 2\delta$). Also, we will consider only cases for which $\alpha > 0.98 $ which corresponds to compounds that present a first order smectic-isotropic 
phase transition according to the McMillan's mean field approach. In this regime, the self-consistent equations exhibit 
two locally stable solutions, corresponding to a smectic phase and a melted center phase which has a null smectic order near the film center\cite{canabarro}. The equilibrium state is determined from the
global minimum of the Helmholtz free-energy.

\begin{figure}[ht!]
\centerline{\includegraphics[width=65mm,clip]{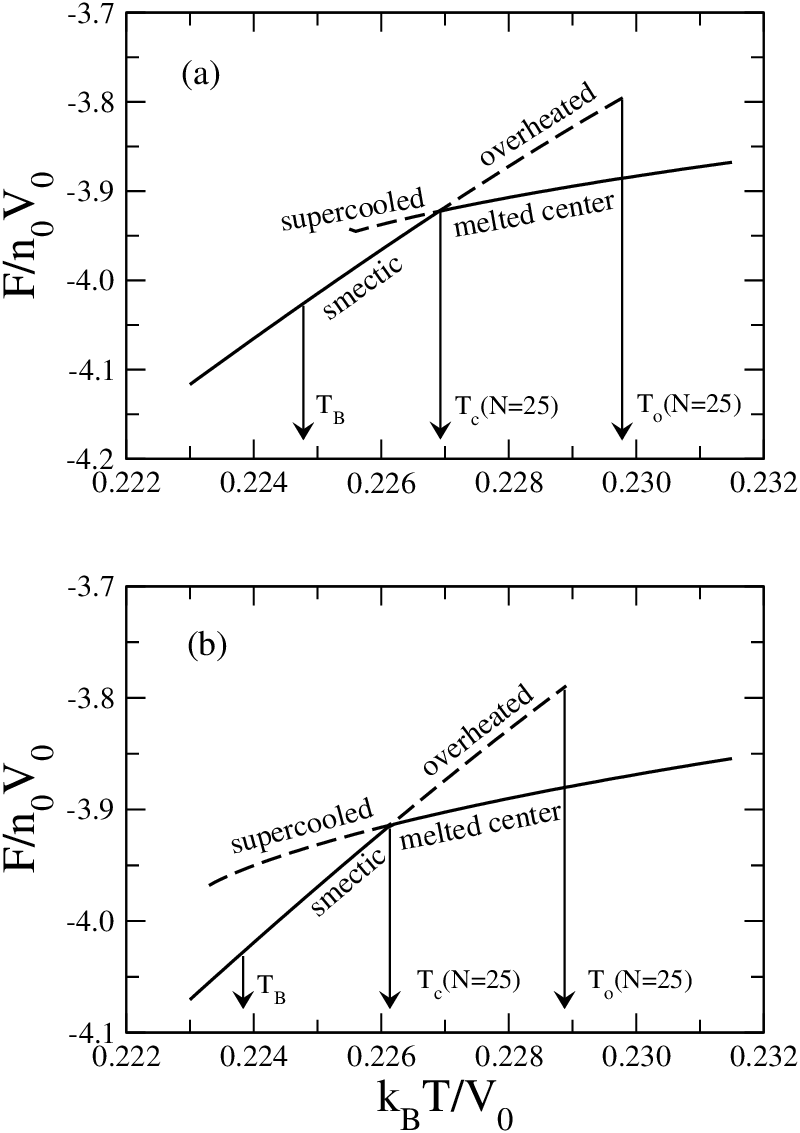}}
\caption{Helmholtz free-energy as a function of temperature
for the solutions of the self-consistent equations which are locally stable.  
The model parameters are $N = 25$, $W_0 = 2.5~V_0$, and  $\alpha = 1.05$. (a)$\sqrt{|\varepsilon_a^*|/V_0}E = 0$
and (b) $\sqrt{|\varepsilon_a^*|/V_0}E = 0.12$. Solid lines represent the global stable solution and
dashed lines correspond to supercooled and overheated states. Notice that the external field reduces
the transition temperature $T_c(N)$ associated with the melting of the central layers, as well as the bulk transition temperature $T_B$ and the limiting temperature $T_o(N)$ above which the overheated phase becomes unstable.} \label{fig1}
\end{figure}

In Fig.1 we present the Helmholtz free-energy as a function of the temperature 
for the locally stable solutions of the self-consistent relations. The model parameters 
used were $N=25$, $W_0 = 2.5 V_0$, and $\alpha = 1.05$, which corresponds to a zero-field
bulk transition temperature of $T_B = 0.22482V_0/k_B$. At zero field, a film with $N$ layers exhibits a transition 
temperature $T_c(N)$ at which the film center melts. For $N=25$, we obtain 
$T_c(N=25) = 0.22695V_0/k_B$, as shown in Fig.1(a). Above $T_c(N)$, the solution with
a non-null smectic order in the center of the film corresponds
to a metastable overheated smectic phase.
The equilibrium solution above $T_c(N)$ is the melted center one. Below $T_c(N)$, a metastable supercooled melted center phase can be 
reached, although the equilibrium solution is the smectic phase. 
Both metastable solutions only appear in a finite temperature range around $T_c(N)$, as expected near a first-order transition.
In the presence of an external electric field, 
we notice that the transition temperature is shifted to a lower value, as shown in Fig.1(b). 
Here, we used $\sqrt{|\varepsilon_a^*|/V_0}E = 0.12$. Further, we notice an enhancement in the 
range of temperatures at which the supercooled isotropic phase is locally stable.

\begin{figure}[ht!]
\centerline{\includegraphics[width=65mm,clip]{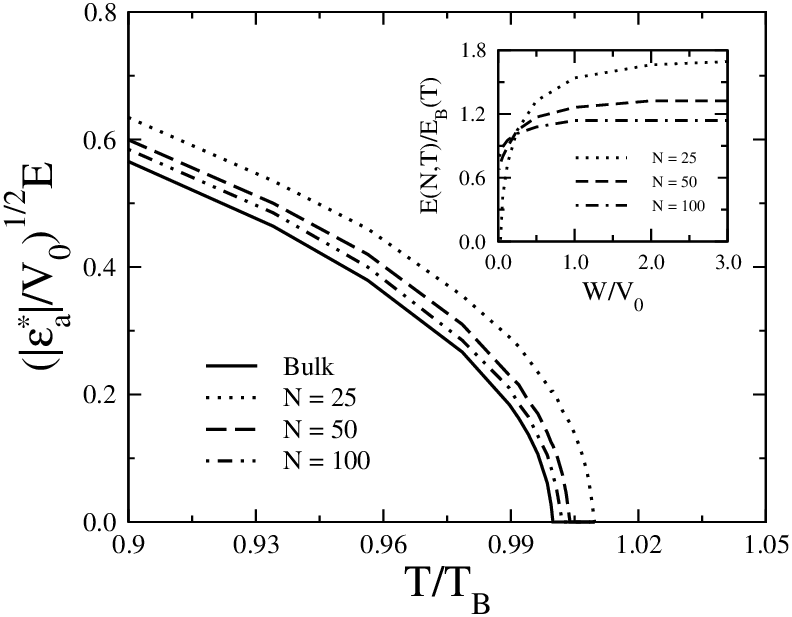}}
\caption{Phase diagram in the $E\times T$ parameter space for $\alpha = 1.05$, $W_0 = 2.5~V_0$, and 
several film thicknesses. In this strong anchoring regime, the transition temperature increases 
as the film thickness reduces. Here $T_B$ is the zero-field bulk transition temperature. The inset
shows the transition field as a function of the surface anchoring for $T = 0.2230~V_0/k_B$ and distinct film thicknesses. The crossing point delimits the weak and strong anchoring regimes.} \label{fig2}
\end{figure}

\begin{figure}[ht!]
\centerline{\includegraphics[width=65mm,clip]{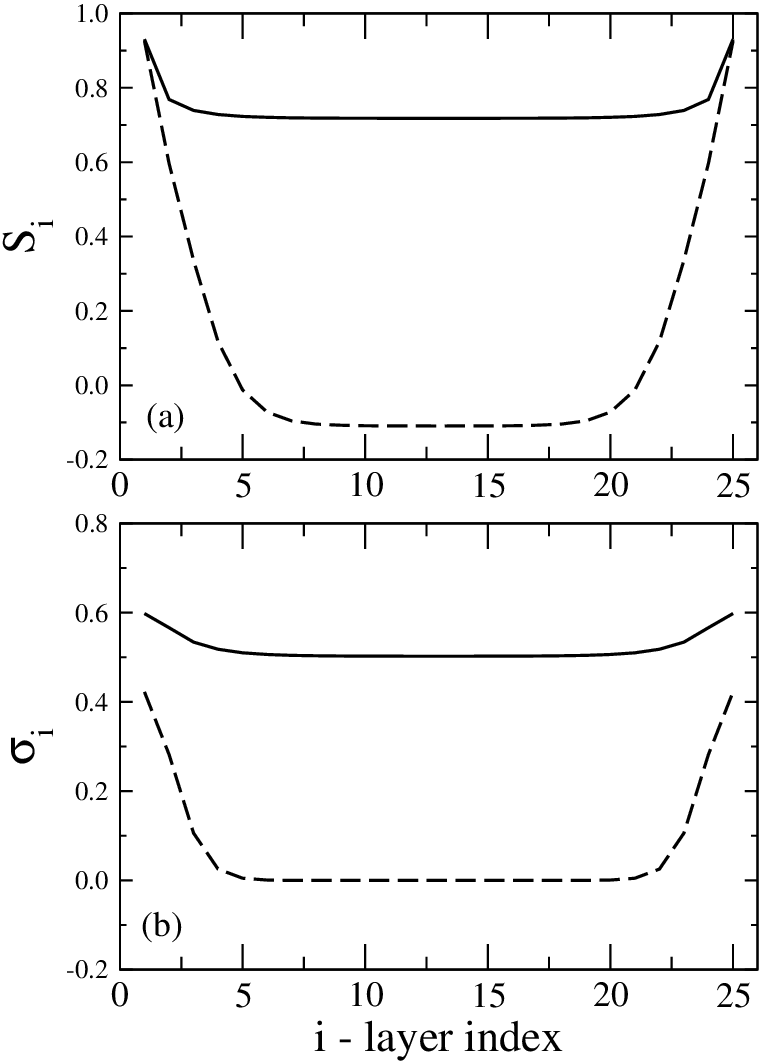}}
\caption{Profiles of the (a) nematic and (b) smectic order parameters
for different external fields: $\sqrt{|\varepsilon_a^*|/V_0}E = 0.270$ (solid lines) and
$\sqrt{|\varepsilon_a^*|/V_0}E = 0.275$ (dashed lines). Here, $T = 0.2230~V_0/k_B$, $N =25$,
and $W_0 = 2.5~V_0$. The zero-field bulk transition temperature is $T_B = 0.22482V_0/k_B$. Notice the field-induced discontinuous transition from the state with finite nematic and smectic order parameters to the melted state at the film center. } \label{fig3}
\end{figure}

The phase diagram showing the field dependence of the transition 
temperature for films with distinct thicknesses is reported in Fig.2, for the
particular case of $W_0 = 2.5~V_0$ and $\alpha = 1.05$. The bulk phase diagram
is also exhibited. We can observe that the transition temperature is higher
for thin films under such strong surface anchoring. Further, the molecular reorientation 
promoted by the external field reduces the transition temperature. The inset shows the transition
field as a function of the surface anchoring for several film thicknesses for 
$T =0.2230~V_0/k_B$, which is below the zero-field bulk transition temperature.  The 
curves crosses roughly at a common point which delimits two anchoring regimes.
Below the crossing point, the transition field increases with the film thickness,
thus leading to the melting of the entire film. On the other hand, the transition field 
is larger for thin films in the strong surface anchoring regime. Therefore, the field-induced 
transition in this regime corresponds to the melting of inner layers, as illustrated 
in Fig.3. In contrast with a null smectic order parameter near the film center, 
the nematic order parameter becomes negative which reflects the field-induced reorientation 
of the molecular alignment \cite{oleg}. The melting of the central layers in the regime of strong
surface anchoring is the typical scenario leading  to the layer thinning phenomenon.

\begin{figure}[ht!]
\centerline{\includegraphics[width=65mm,clip]{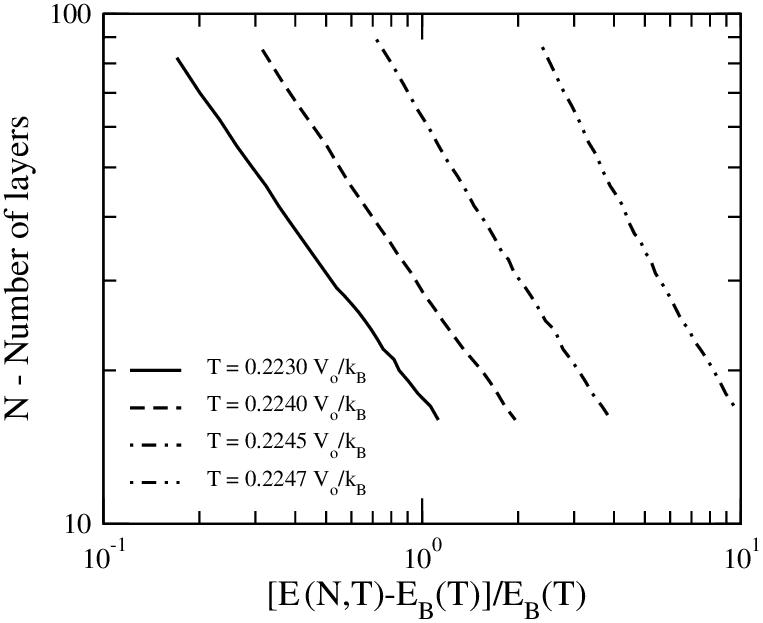}}
\caption{External field dependence of the film thickness
for different values of the temperature. We used $\alpha = 1.05$ and $W_0 = 2.5~V_0$.
$E_B(T)$ is the bulk transition field at temperature $T$. In the thickness range shown, the 
thinning transition follows closely a power law 
dependence with the reduced external field $E(T)-E_B(T)$.} \label{fig4}
\end{figure}

\begin{figure}[ht!]
\centerline{\includegraphics[width=65mm,clip]{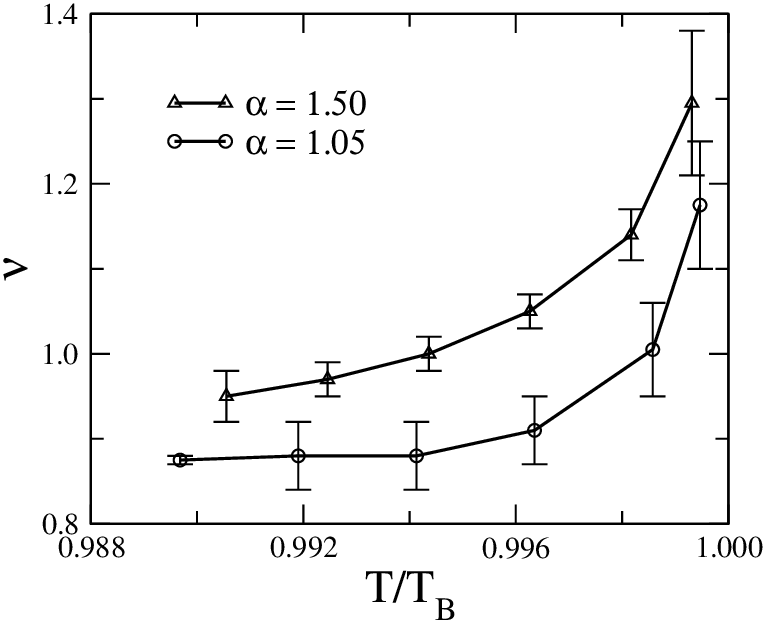}}
\caption{Temperature dependence of the characteristic  exponent $\nu$ of
the field-induced layer thinning transition for $W_0 = 2.5~V_0$. The error bars account for the small variability
of the exponent along the transition line in the range of film thicknesses shown in Fig.~4. The exponent $\nu$
increases monotonically as the bulk transition temperature is approached.} \label{fig5}
\end{figure}

In the absence of an external field, the layer thinning transition corresponds
to a stepwise power-law reduction of the film thickness as the temperature is raised
above the bulk transition temperature \cite{stoebe,canabarro}. The above
results indicate that a similar layer thinning transition can take place below
the bulk transition temperature with the film thickness controlled by an external
field. In Fig. 4 we exhibit the field dependence of the film thickness for
distinct values of the temperature in the regime of strong surface anchoring. 
One can notice a continuous thinning of the film thickness as 
the external field exceeds the bulk transition field.
Such behavior can be reasonably described by a power law $N(E)\propto[E(T)-E_B(T)]^{-\nu}$,
where $E_B(T)$ is the bulk transition field.
The different slopes of the curves point to a slight temperature dependence of the exponent $\nu$.
Fig. 5 shows that the power-law exponent $\nu$ is smaller than unity far from 
the bulk transition temperature. 
These values are of the same order of magnitude of those
observed in temperature-induced layer thinning transitions \cite{canabarro}.
However, $\nu$ becomes larger close to the bulk transition temperature. Further,
we notice that power-law exponent depends on the parameter $\alpha$, which
is related to the length of the alkyl chains of the liquid crystal molecules.
The reduction of the film thickness is faster in liquid crystal compounds 
with longer alkyl chains.

In summary, we demonstrated that a layer thinning transition can be induced 
by an external electric field in free standing smectic films, under strong surface anchoring and 
below the bulk transition temperature. We considered a system with a negative dielectric 
anisotropy on which an external field perpendicular to the layer plane 
can promote a Fr\'eedericksz transition in the center of the film. The reorientation of 
the molecular alignment is accomplished by the melting of the inner smectic layers
thus leading to the layer thinning transition. Our results showed that the number of smectic
layers decays monotonically with the external field. In the range of film thicknesses investigated,
the reduction in the number of layers is well described by a power-law
$N(E)\propto[E(T)-E_B(T)]^{-\nu}$ above the field $E_B(T)$ that is able to reorient bulk samples
at temperature $T$. The effective exponent $\nu$ increases as the temperature approaches to the
bulk transition temperature. Further, our results indicated that the effective exponent
$\nu$ is larger in compounds with longer alkyl chains. Considering the typical liquid crystal
physical parameters, the presently predicted field-induced layer thinning transition can be 
experimentally observed for electric fields of the order of $10^5$~V/cm, which is well within 
achievable values \cite{lelidis,rosenblatt}. Efforts in this direction would bring 
valuable information concerning the interplay of field, surface and finite-size effects
in the phase transitions depicted by free-standing smectic films.

\begin{acknowledgments}
We would like to thank CAPES, CNPq, and FINEP (Brazilian
Research Agencies) as well as FAPEAL (Alagoas State
Research Agency)for partial financial support.
\end{acknowledgments}


\begin{thebibliography}{40}

\bibitem{bahr}C. Bahr, Phys. Rev. Lett. \textbf{99}, 057801 (2007).
%
\bibitem{chao}C. Y. Chao, C.R. Lo, P.J. Wu, T.C. Pan, M. Veum, C.C. Huang, V. Surendranath, and J.T. Ho, Phys. Rev. Lett. \textbf{88}, 085507 (2002).
%
\bibitem{barbero}G. Barbero and L.R. Evangelista, Phys. Rev. E \textbf{65}, 031708 (2002).
%
\bibitem{sprunt}T. Ostapenko, D.B. Wiant, S.N. Sprunt, A. J\'akli, and J.T. Gleeson, Phys. Rev. Lett. \textbf{101}, 247801 (2008).
%
\bibitem{lelidis}I. Lelidis, Phys. Rev. Lett. \textbf{86}, 1267 (2001).
%
\bibitem{rosenblatt}B. Wen and C. Rosenblatt, Phys. Rev. Lett. \textbf{89}, 195505 (2002).
%
\bibitem{degennes} P.G. de Gennes and J. Prost, \textit{The Physics of Liquid Crystals}(Clarendon Press, Oxford, 1993).
%
\bibitem{bahr1}C. Bahr, Int. J. Mod. Phys. B \textbf{8}, 3051 (1994).
%
\bibitem{stoebe}T. Stoebe, P. Mach, and C.C. Huang, Phys. Rev. Lett. \textbf{73}, 1384 (1994).
%
\bibitem{mol}E.A. Mol, G.C.L. Wong, J.M. Petit, F. Rieutord, and W.H. de Jeu, Physica B \textbf{248}, 191 (1998).
%
\bibitem{demikhov}E.I. Demikhov, V.K. Dolganov, and K.P. Meletov, Phys. Rev. E \textbf{52}, R1285 (1995).
%
\bibitem{jin}A.J. Jin, M. Veum, T. Stoebe, C.F. Chou, J.T. Ho, S.W. Hui,
V. Surendranath, and C.C. Huang, Phys. Rev. E \textbf{53}, 3639 (1996). 
%
\bibitem{johnson}P.M. Johnson, P. Mach, E.D. Wedell, F. Lintgen, M. Neubert, and C.C. Huang, Phys. Rev. E \textbf{55}, 4386 (1997).
%
\bibitem{pankratz}S. Pankratz, P.M. Johnson, H.T. Nguyen, and C.C. Huang, Phys. Rev. E \textbf{58}, R2721 (1998).
%
\bibitem{picano}F. Picano, P. Oswald, and E. Kats, Phys. Rev. E \textbf{63}, 021705 (2001).
%
\bibitem{mirantsev}L.V. Mirantsev, Phys. Lett. A \textbf{205}, 412 (1995).
%
\bibitem{sullivan}D.E. Sullivan and A.N. Shalaginov, Phys. Rev. E \textbf{70}, 011707 (2004).
%
\bibitem{canabarro}A.A. Canabarro, I.N. de Oliveira, and M.L. Lyra, Phys. Rev. E \textbf{77}, 011704 (2008).
%
\bibitem{lelidis1}I. Lelidis and G. Durand, Phys. Rev. Lett. \textbf{73}, 672 (1994).
%
\bibitem{dolganov2}F. Bougrioua, P. Cluzeau, P. Dolganov, G. Joly, H. T. Nguyen, and V. Dolganov, Phys. Rev. Lett.
\textbf{95}, 027802 (2005).
%
\bibitem{mirantsev2}L.V. Mirantsev, Phys. Rev. E \textbf{55}, 4816 (1997).
%
\bibitem{mcmillan}W.L. McMillan, Phys. Rev. A \textbf{4}, 1238 (1971).
%
\bibitem{selinger}J.V. Selinger and D.R. Nelson, Phys. Rev. A \textbf{37}, 1736 (1988).
%
\bibitem{oleg} Z. Li and O.D. Lavrentovich, Phys. Rev. Lett. \textbf{73}, 280 (1994).
%
\end{thebibliography}
\end{document}